\documentclass[twocolumn]{aastex631}
\hypersetup{linkcolor=red,citecolor=blue,filecolor=cyan,urlcolor=magenta}

\usepackage{CJK}
\usepackage{amsmath} 
\usepackage{bm}      
\usepackage{lmodern}
\usepackage{amssymb}
\usepackage{physics} 
\usepackage{pgfplots}  

\shorttitle{Temperature Separation in a Vortex Tube}
\shortauthors{Chen Haibin et al.}

\graphicspath{{./}{figures/}}

\begin{document}
\begin{CJK*}{UTF8}{gbsn}
\title{Temperature Separation in a Vortex Tube and convection driven by vorticity}

\author[0000-0002-5500-3634]{Haibin Chen (陈海彬)}
\correspondingauthor{Haibin Chen}
\email{chenhb3@mail2.sysu.edu.cn}

\author[0000-0003-0264-4363]{Rong Wu (吴蓉)}

\begin{abstract}


Why the temperature gradient in the vortex tube deviates significantly from the adiabatic gradient is a very important but unresolved issue in vortex tube research. In compressible fluids, the vorticity of a fluid parcel that is different from its environment can affect its own pressure or density, just like temperature. Gases near the wall have very high vorticity, and when they detach from the wall, their higher vorticity compared to the environment makes their density lower than the environment. Driven by vorticity, some fluid parcels undergo multiple collisions near the axis, achieving temperature separation.

\end{abstract}




\section{Introduction}

The phenomenon of temperature separation in a swirling vortex flow was discovered in the 1930s by Georges J. Ranque. Ranque proposed that compression and expansion effects are the main reasons for the temperature separation in the tube \citep{ranque1933experiments}\citep{joseph1934method}. Later, the geometrical parameters and performance optimisation of the tube were investigated by Hilsch \citep{hilsch1947use}. Due to the complexity of the flow structure in the tube, none of the aforementioned factors is proven to be the real reason for energy separation in the vortex tube \citep{XUE20101367}.


A new research\citep{2022arXiv220711990C}\citep{2022arXiv221108113C}\citep{2024arXiv240110105C} points out that rotation or vorticity can play the role of temperature in a rotating system, promoting or suppressing convection. This theory may also apply to vortex tubes.

\section{Convection criterion of rotating fluid}

%
%
%

In the vortex tubes, the rotational kinetic energy of fluid parcels may approach or even exceed their thermal energy. This leads to convective flows being influenced by the rotation of the fluid parcels.

The thermal energy of a fluid parcel is given by
\begin{equation}
E_T = \frac{\rho V R_m}{M_m} T
\end{equation}
while its rotational kinetic energy can be expressed as
\begin{equation}
E_r \propto \rho V \Omega^2 l^2
\end{equation}
The ratio between these two energies can be represented by a dimensionless parameter
\begin{equation}
Tr = \frac{\frac{R_m}{M_m} T}{\Omega^2 l^2}
\end{equation}
This dimensionless parameter can also be written as the square of the ratio between the speed of sound wave propagation $v_s$ and the propagation speed of inertial waves $v_\Omega$
\begin{equation}
Tr \propto \frac{v_s^2}{v_\Omega^2}
\end{equation}
When $Tr \gg 1$, thermal energy dominates the motion of the fluid parcel. As $Tr$ approaches 1, rotation can directly influence the motion of the fluid parcel.

According to calculations, the $Tr$ value for solar granules can be as high as $10^6$, indicating that their motion is primarily driven by thermal energy and rotation cannot directly affect it. Therefore, the claim made in the article from \citet{2022arXiv220711990C}\citet{2022arXiv221108113C}\citet{2024arXiv240110105C}  that the size of granules is directly influenced by rotation is incorrect. For rotation to have an impact on solar thermal convection, multiple amplifications may be necessary. However, the study of rotating convection in the article provides insights for vortex tubes.

In vortex tubes, the maximum value of $Tr$ can approach 1, suggesting that rotation can directly influence convection.

\section{Convection criterion of rotating fluid}
The effects of rotation on convection are scattered throughout several papers on solar physics. Before applying them to vortex tubes, we will introduce and summarize them.

\subsection{Schwarzschild convection criterion}

Under normal conditions, we can use the Schwarzschild convection criterion\citep{1950ApJ...111..351R} to determine whether convection will occur spontaneously:
\begin{equation}
\left|\frac{dT}{dR}\right|_{\text{rd}} > \left|\frac{dT}{dR}\right|_{\text{ad}}
\end{equation}
Where $\left(\frac{dT}{dR}\right)_{\text{rd}}$ is the real temperature gradient, and the subscript can be dropped during calculation. $\left(\frac{dT}{dR}\right)_{\text{ad}}$ is the adiabatic temperature gradient, which represents the temperature change caused by the expansion or compression of a fluid parcel in an adiabatic state as it moves from one region to another.

The pressure equilibrium between the fluid parcel and its environment is established relatively quickly, while the thermal equilibrium is slower. Under perturbed conditions, the fluid parcel can be considered adiabatic. If the temperature gradient of the environment is not equal to the adiabatic gradient, after being perturbed by a displacement in the direction of gravity or opposite to it, the temperature of the fluid parcel will differ from that of the environment, and its density will also differ, resulting in a mismatch between buoyancy and gravity. When the convection criterion is satisfied, the combined force of gravity and buoyancy will point away from the equilibrium position, placing the fluid in an unstable state where convection will occur spontaneously. When the temperature gradient is equal to the adiabatic gradient, it has no effect on the motion of the fluid. When the temperature gradient is less than the adiabatic gradient, motion in the direction of gravity will be restricted, resulting in oscillations around the equilibrium position.

For convenient calculation, the adiabatic temperature gradient can be expressed using the real pressure gradient or the real density gradient:
\begin{equation}
\left(\frac{dT}{dR}\right)_{\text{ad}} = (\gamma - 1)T \frac{d\rho}{\rho dR}
\end{equation}
In this paper, the real density gradient is chosen to represent the adiabatic temperature gradient, which avoids interference from pressure generated by other factors.

The convection criterion generally assumes that the signs of the temperature gradient, density gradient, and pressure gradient are the same. In this case, adding absolute value symbols eliminates the need to discuss their signs, making it more convenient to use. However, in special cases, classification and discussion may be required.

\subsection{Equivalent Temperature of Rotation or Vorticity}

In a vortex tube, the rotational speed of the fluid is very high, and the pressure generated by the rotation can even approach that generated by temperature. This additional pressure can change the density of the convective fluid parcel and affect the buoyancy. Therefore, the influence of rotation must be considered in the convection within the vortex tube.

As an example, consider a cylindrical fluid parcel rotating about its own axis of symmetry at a speed of $\Omega$, with a radius of $l$ and a height of $2l$, and a fluid density of $\rho$. The average increase in surface pressure of the fluid cylinder due to rotation is given by:
\begin{equation}
\bar{p}_{\Omega} = \frac{1}{6} \rho \Omega^2 l^2
\end{equation}
During the isotropic compression (expansion) of the fluid parcel, this pressure causes an increase (decrease) in its rotational energy, similar to the relationship between gas pressure and density with three degrees of freedom. Therefore, the influence of rotation can also be described using an equivalent rotational temperature, which facilitates the incorporation of the effects of rotational energy into the convection criterion. Following the example of the gas state equation, we let:
\begin{equation}
\bar{p}_{\Omega} = \rho \frac{R_m}{M_m} T_{\Omega}
\end{equation}
Where $R_m$ is the gas constant, $M_m$ is the molar mass of the gas, and $T_{\Omega}$ is the newly defined equivalent rotational temperature. Then $T_{\Omega} = \frac{M_m}{6R_m} \Omega^2 l^2$. For different shapes and axes of rotation, such as spheres, squares, or cylinders with axes of rotation not aligned with their symmetry axes, the coefficients of the additional pressure and equivalent temperature generated by rotation will be different. Let:
\begin{equation}
T_{\Omega} = k_{\Omega} \frac{M_m}{R_m} \Omega^2 l^2
\end{equation}
Where the value of $k_{\Omega}$ depends on the shape and axis of rotation, and can be calculated based on the change in rotational energy during isotropic expansion.

When studying vortex tubes, it is more convenient to use vorticity, so the equivalent vorticity temperature can be used instead:
\begin{equation}
T_{\omega} = k_{\omega} \frac{M_m}{R_m} \omega^2 l^2
\end{equation}
For the same fluid parcel, it is generally true that $\omega = 2\Omega$, and by letting $k_{\omega} = 4k_{\Omega}$, we can ensure that $T_{\omega} = T_{\Omega}$.

The concept of $\omega^2$ in the formula is the same as that in the pseudo-enstrophy $\omega^2/2$, both representing the magnitude of the rotation speed of the fluid parcel, without considering the direction. Here, $\omega=|\vec{\omega}|$.

The deformation of the fluid parcel can also affect its pressure distribution, but in vortex tubes, the vorticity is generally much larger than the deformation, so the influence of deformation can be ignored.

When the life time of the fluid parcel is much shorter than its rotation period, its deformation is statistically isotropic. Turbulence and inertial oscillations can cause the deformation of the fluid parcel to deviate from isotropy. There are more detailed theoretical studies and corresponding observational phenomena in solar physics\citep{2024arXiv240110105C}, but this situation is temporarily ignored in this paper.

\subsection{Vorticity Gradient-Driven Convection}

The rotation or vorticity of fluid parcels can independently drive or suppress convection, just like the temperature. Assuming the temperature gradient equals the adiabatic gradient, in the absence of rotational effects, buoyancy and gravity are always equal during the motion of fluid parcels, leaving them in a state of neutral equilibrium. If the influence of rotation on density is non-negligible, rotation will alter the force state of the fluid parcels, promoting or suppressing convection.

Analogous to the Schwarzschild criterion for convection, the convective criterion determined by rotation for isotropically expanding or compressing fluid parcels is given by
\begin{equation}
\left|\frac{dT_{\omega}}{dR}\right|_{\text{rd}} > \left|\frac{dT_{\omega}}{dR}\right|_{\text{ad}}
\end{equation}
The adiabatic gradient of the vorticity-equivalent temperature can be expressed in terms of the density gradient as
\begin{equation}
\left(\frac{dT_{\omega}}{dR}\right)_{\text{ad}} = \frac{2}{3} T_{\omega} \frac{d\rho}{\rho dR}
\end{equation}
Since the size of fluid parcels also affects the vorticity-equivalent temperature, the vorticity-equivalent temperature gradient $\left( \frac{dT_{\omega}}{dR} \right)_{rd}$ needs to be expressed simultaneously in terms of the vorticity gradient $\frac{d\omega}{\omega dR}$ and the change in size of the fluid parcels during expansion $\frac{d l}{l dR}$, where $l$ can be expressed in terms of the density gradient, i.e.,
\begin{equation}
\left(\frac{dT_{\omega}}{dR}\right)_{\text{rd}} = 2T_{\omega} \frac{d\omega}{\omega dR} - \frac{2}{3} T_{\omega} \frac{d\rho}{\rho dR}
\end{equation}
After rearranging, the convective criterion obtained is
\begin{equation}
\left|\frac{d\omega}{\omega dR}\right| > \left|\frac{2}{3} \frac{d\rho}{\rho dR}\right|
\end{equation}
In a fluid where the temperature gradient satisfies the adiabatic gradient, the temperature gradient has no effect on the fluid's motion. When the vorticity gradient satisfies the convective criterion, it drives convection. When the vorticity gradient does not satisfy the convective criterion, it has no effect on the fluid's motion, or motion along the direction of gravity or centrifugal force becomes restricted, resulting in vibrations near the equilibrium position.

\subsection{Convective Criterion Determined Jointly by Vorticity Gradient and Temperature Gradient}

When both the temperature gradient and the vorticity gradient satisfy the convective criterion, they jointly drive convection. When they both fail to satisfy the convective criterion, they have no effect on or suppress the fluid's motion, which is obvious. When only one of them satisfies the convective criterion while the other does not, the situation becomes complex, and some interesting phenomena emerge. Due to the larger size of fluid parcels correlating with higher vorticity-equivalent temperatures, this results in larger flows being dominated by the vorticity gradient, while smaller fluid parcels are dominated by the temperature gradient, leading to a size-dependent convective criterion.

For isotropically compressing or expanding fluid parcels, considering both the temperature gradient and the vorticity gradient, the convective criterion is given by
\begin{equation}
\left|\frac{dT + dT_{\omega}}{dR}\right|_{\text{rd}} > \left|\frac{dT + dT_{\omega}}{dR}\right|_{\text{ad}}
\end{equation}
Due to the complex nature within vortex tubes, the simple criterion with absolute values is no longer applicable, and the absolute value signs need to be removed. If $R$ is defined as the distance from the target point to the axis of the vortex tube, then the positive direction of $R$ is the same as the centrifugal force direction, and the pressure gradient and density gradient are positive, allowing the direct removal of the absolute value signs. Rearranging the convective criterion gives
\begin{equation}
\frac{dT}{dR} + 2T_{\omega} \frac{d\omega}{\omega dR} - \frac{2}{3} T_{\omega} \frac{d\rho}{\rho dR} > (\gamma - 1)T \frac{d\rho}{\rho dR} + \frac{2}{3} T_{\omega} \frac{d\rho}{\rho dR}
\end{equation}
Substituting $T_{\omega}$ into the equation yields a size-dependent convective criterion. Letting the critical size be $l_{\text{ad}}$, we have
\begin{equation}  
	l_{ad}^2 = \frac{-T\left(\frac{dT}{TdR} - (\gamma-1)\frac{d\rho}{\rho dR}\right)}{\frac{2k_{\omega}M_m}{R_m}\omega^2\left(\frac{d\omega}{\omega dR} - \frac{2}{3}\frac{d\rho}{\rho dR}\right)}  
\end{equation}

Based on the vorticity distribution and temperature distribution, there are two scenarios. If the vorticity gradient satisfies the convective criterion while the temperature gradient does not, i.e., $\frac{dT}{TdR} < (\gamma - 1) \frac{d\rho}{\rho dR}$ and $\frac{d\omega}{\omega dR} > \frac{2}{3} \frac{d\rho}{\rho dR}$, convection is driven by the vorticity gradient, and the size-dependent form of the convective criterion is $l > l_{\text{ad}}$. If the temperature gradient satisfies the convective criterion while the vorticity gradient does not, i.e., $\frac{dT}{TdR} > \left| \gamma - 1 \right| \frac{d\rho}{\rho dR}$ and $\frac{d\omega}{\omega dR} < \frac{2}{3} \frac{d\rho}{\rho dR}$, convection is driven by the temperature gradient, and the size-dependent form of the convective criterion is $l < l_{\text{ad}}$.

The size-dependent convective criterion can address some issues related to thermal convection in rotating systems. In some regions of vortex tubes, the temperature gradient can be significantly larger than the adiabatic gradient, which is attributed to the low efficiency of heat transport caused by the absence of larger fluid parcels. Simultaneously, vorticity-driven motion is the core mechanism behind temperature separation in vortex tubes.

\section{Motion in Vortex Tubes}

\subsection{Convection Criteria in Vortex Tubes}

In a vortex tube, two regions can be distinguished based on the direction of vorticity: an inner region where the average vorticity aligns with the rotation direction, and an outer region where it opposes it. In the inner region, the vorticity gradient $\frac{d\omega}{\omega dR}$ is negative\citep{eiamsa2007numerical}, indicating that this region does not satisfy the convection criteria for vorticity. Convection in this region is suppressed by the vorticity, and even if a portion of this region meets the convection criteria for temperature, the size of the resulting fluid parcels will be smaller than $l_{ad}$, which may have a relatively small value, providing good insulation for the inner region.

\begin{figure*}[htbp]  
	\centering  
	\includegraphics[width=\textwidth]{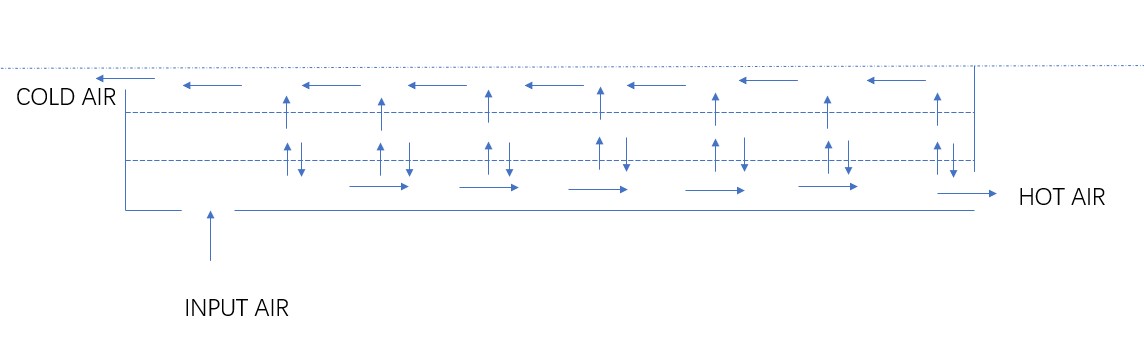}  
	\caption{mainly gas flow in a vortex tube}  
	\label{fig:flow}  
\end{figure*}

In the outer region of the vortex tube, the average vorticity opposes the rotation direction, and the $\frac{d\omega}{\omega dR}$ is positive\citep{behera2008numerical}. Due to strong friction with the wall, the value of $\frac{d\omega}{\omega dR}$ is very large. During the operation of the vortex tube, it can be considered to satisfy the convection criteria for vorticity, and under normal circumstances, the value of $\frac{d\omega}{\omega dR}$ is significantly greater than $\frac{2}{3} \frac{d\rho}{\rho dR}$. This results in a strong convective trend in the outer region, which accelerates radially along the vortex tube when perturbed. This results in the static temperature gradient in the outer region being very close to the adiabatic gradient\citep{bruun1969experimental}.

\subsection{Fluid Parcels Jumping from Near the Wall to Near the Axis}

Fluid parcels near the wall have very high vorticity. As they move radially inward, they can cross the boundary between the inner and outer regions where the vorticity is zero. At this point, the vorticity of the fluid parcel is opposite to that of the environment but higher in magnitude. The additional pressure from the vorticity can still keep the density of the fluid parcel lower than that of the environment, allowing it to continue moving towards the axis under the influence of buoyancy. During this process, expansion lowers the temperature and vorticity of the fluid parcel, and some high-vorticity fluid parcels can drive themselves to reach the axis of the vortex tube.

Due to the suppressed flow in the inner region, the only fluid with radial velocity in this region, apart from the overall radial flow, comes from fluid parcels near the wall of the outer region. The vorticity of these moving fluid parcels is opposite to that of the inner region, a distinct feature that may be observable in experiments.

\subsection{Angular Momentum Screening Mechanism and Thermal-Dynamic Separation}

As fluid parcels move inward from near the wall of the vortex tube, their swirl velocity increases as they approach the axis, due to the conservation of angular momentum. This increases the centrifugal force on the fluid parcels, preventing them from moving further towards the axis.

We hypothesize that within the inner layer, moving fluid parcels impact the surrounding fluid and exchange energy and angular momentum with it. After exchanging angular momentum, the fluid parcels experience a reduction in angular momentum. When the high swirl speed-induced Coriolis force is eliminated, the fluid parcels continue to move inward. During this process, some fluid parcels may fragment. If the resulting fragments are too small to provide sufficient buoyancy, they will either remain in place or move outward under the influence of temperature gradients.

As fluid parcels traverse the entire vortex tube from near the wall to near the axis, their rotational speed decreases significantly, resulting in a substantial loss of angular momentum. This may be the core mechanism behind the thermal-dynamic separation observed in vortex tubes.

After fluid parcels from the outer region pass through the inner region, there is an increase in both energy and angular momentum, which can cause the surrounding fluid to vibrate. This vibration may correspond to the observed relationship between sound and temperature separation in vortex tubes\citep{kurosaka1982acoustic}\citep{frohlingsdorf1999numerical}. The kinetic energy of the fluid may be converted into thermal energy during these vibrations.

The frequency of this vibration can be calculated using vorticity gradients, temperature gradients, and density gradients.

%
%
%
%
%
%
%
%
%
%
%
%
%
%

\section{Temperature Separation in Vortex Tubes}

Integrating previous theories and research, we can speculate on the flow within vortex tubes. We propose that the primary flow within vortex tubes can be divided into three regions:

1. The outer region, where gas enters this region from the nozzle and flows along the wall of the vortex tube toward the hot outlet. During this process, kinetic energy is continually dissipated, and the temperature gradually increases, reaching a maximum at the hot end. This process dominates the axial temperature of the vortex tube. As the gas moves from the nozzle to the hot end, some of the high-vorticity gas near the wall crosses the entire vortex tube, losing kinetic energy and angular momentum as it approaches the axis. In areas closer to the nozzle, the vorticity near the wall is higher, providing stronger driving force, enabling more gas to reach near the axis from the wall. Fluids in regions closer to the hot end have more difficulty reaching near the axis.

2. The cold gas collection zone in the inner region, located near the axis, collects gas arriving from near the vortex tube wall and exits through the cold end outlet. Since a higher proportion of fluid from regions closer to the nozzle reaches this area, the temperature at the cold end outlet is lower. Simultaneously, the gas near the axis also exhibits lower dynamic temperature.

3. The insulating layer between the inner and outer regions, which isolates the outer region from the cold gas collection zone. Thermal convection in this area is suppressed by vorticity, resulting in relatively good thermal insulation effects. Additionally, centrifugal force imbalances can induce secondary flows, which may have some influence on the insulation performance of this region. Fluid parcels that break up in this region can replenish the fluid in this area, while most of the fluid recirculates back to the outer region. The vibration of fluid in this region dissipates a significant amount of kinetic energy and converts it into thermal energy. The high-temperature gas recirculating back to the outer region may be an important mechanism for temperature rise in the outer region.

The division of areas and the mainly gas flow are shown in the figure\ref{fig:flow}.

%
%
%
%
%

\section{Conclusion}

This paper has reorganized the theory of studying rotating convection in solar physics and applied it to the investigation of vortex tubes. According to this theory, convection in the inner region of the vortex tube is suppressed. Even if the temperature gradient is higher than the adiabatic gradient, the size of fluid parcels is limited, similar to solar granulation. Fluid parcels near the wall in the outer region possess very high vorticity. When they move to other regions, the increased pressure due to high vorticity makes their density lower than the surrounding environment. Buoyancy drives the fluid parcels to move inward, and some fluid parcels can traverse the entire vortex tube to reach near the axis, then flow out from the cold end along the axial direction. During this process, fluid parcels exchange kinetic energy and angular momentum with the surrounding fluid, reducing their own kinetic energy and exhibiting lower temperatures. The gas flowing out from the hot end carries a large amount of kinetic energy and the thermal energy converted from kinetic energy, resulting in a higher temperature. 

The new theory presents a simple and clear model for the temperature operation of vortex tubes, representing a significant breakthrough in vortex tube research.

%
%
%

\section*{acknowledgments}
The data underlying this article are available in the article and in its online supplementary material.
This paper was translated by AI Wenxinyiyan.

\bibliography{reference}{}
\bibliographystyle{aasjournal}

\end{CJK*}
\end{document}